\documentclass[10pt]{article}

\usepackage{amsmath}
\usepackage{amssymb}

\usepackage{graphicx}

\usepackage{cite}

\usepackage{color} 

\usepackage[labelfont=bf,labelsep=period,justification=raggedright]{caption}

\makeatletter
\renewcommand{\@biblabel}[1]{\quad#1.}
\makeatother

\date{}

\newcommand{\ben}{\begin{eqnarray}}
\newcommand{\een}{\end{eqnarray}}
\newcommand{\be}{\begin{equation}}
\newcommand{\ee}{\end{equation}}

\begin{document}

\begin{flushleft}
{\Large
\textbf{Spatial and topological organization of DNA chains induced by gene co-localization}
}
\\
Ivan Junier$^{1,2}$,
Olivier Martin$^{3}$$^\ddag$, 
Fran\c{c}ois K\'ep\`es$^{1}$$^\ddag$
\\
$^1$ Epigenomics Project, Genopole, CNRS UPS 3201, UniverSud Paris, University of Evry, Genopole Campus 1 - Genavenir 6, 5 rue Henri Desbru\`eres, F-91030 Evry cedex, France
\\
$^2$ Institut des Syst\`emes Complexes Paris \^Ile-de-France, 57-59 rue Lhomond, F-75005, Paris, France
\\
$^3$ Univ Paris-Sud, UMR 8626 LPTMS, F-91405, Orsay, France,
Univ Paris-Sud, UMR 0320 / UMR 8120 G\'en\'etique V\'eg\'etale, 
F-91190, Gif/Yvette, France
\\
$\ddag$ Equal contribution of these authors.
\\
$\ast$ E-mail: francois.kepes@epigenomique.genopole.fr
\end{flushleft}

\section*{Abstract}
Transcriptional activity has been shown to relate to the organization of chromosomes in the eukaryotic nucleus and in the bacterial nucleoid.
In particular, highly transcribed genes, RNA polymerases and transcription factors gather into discrete spatial foci called 
transcription factories. However, the mechanisms underlying  the formation of these foci and the resulting topological 
order of the chromosome remain to be elucidated. Here we consider a thermodynamic framework based on a worm-like chain 
model of chromosomes where sparse designated sites along the DNA are able to interact whenever they are spatially close-by. This is motivated by recurrent evidence that there exists physical interactions between 
genes that operate together. Three important results come out of this simple framework. First, 
the resulting formation of transcription foci can be viewed as a micro-phase separation of the interacting sites from the
rest of the DNA. In this respect, a thermodynamic analysis suggests transcription factors to be appropriate candidates for mediating the physical interactions between genes. Next, numerical simulations of the polymer reveal a rich variety of phases that are associated with different
topological orderings, each providing a way to increase the local concentrations of the interacting sites. Finally, 
the numerical results show that both one-dimensional clustering and periodic location of the binding sites along the DNA, which have been observed in several organisms, make the spatial co-localization of multiple families of genes particularly efficient. 

\section*{Author Summary}

The good operation of cells relies on a coordination between chromosome structure and genetic regulation which is yet to be understood.
This can be seen in particular from the transcription machinery: in some eukaryotes and bacteria, transcription of highly active genes 
occurs within discrete foci called transcription factories, where RNA polymerases, transcription factors and their target genes co-localize. 
The mechanisms underlying  the formation of these foci and the resulting topological structure of the chromosome remain to be elucidated.
Here, we propose a thermodynamic framework based on a polymer description of DNA in which genes effectively interact through attractive forces in physical space. 
The formation of transcription foci then corresponds to a self-organizing process whereby the interacting genes and the non-interacting DNA form two phases 
that tend to separate. Numerical simulations of the model unveil a rich zoology of the topological ordering of DNA around the foci and show that regularities 
in the positions of the interacting genes make the spatial co-localization of multiple families of genes particularly efficient. Experimental test of the predictions of our model 
should shed new light on the relation between transcriptional regulation and cellular conformations of chromosomes.

\section*{Introduction}

The proper genome-wide coordination of gene expression has been shown to be
linked to the spatial organization of genes within the cell
\cite{Cremer:2006tn,Sexton:2007yt}. This can be seen in particular 
from the transcription machinery: in some 
eukaryotes \cite{Jackson:1993ye,Wansink:1993gf} and
bacteria~\cite{Cabrera:2003fu}, transcription of highly active genes 
occurs within discrete foci called transcription factories, where RNA
polymerases, transcription factors (TFs) and their
target genes co-localize. In eukaryotes, genes that are co-localized
in the same nuclear area are thought to 
participate to the same developmental function \cite{Sexton:2007yt}. 
Accordingly, one-dimensionally distant genes, {\it i.e.} genes that are far apart along the DNA, 
participating in
the same cellular function are expected to co-localize in the three-dimensional
cellular space during periods of active transcription, as
has been shown for generally active 
genes \cite{Osborne:2004pi,Simonis:2006dk}. 

It has
been argued that the associated higher concentrations of certain
molecular species allow for more efficient 
transcription regulation~\cite{Muller-Hill:1998db}, 
just as having transcriptional factories allows for more rapid recycling
of the molecular components of the RNA polymerase complex; both of these
aspects justify a posteriori conformational organizations of the DNA to produce
co-localization phenomena.  
On the experimental side, the 3-dimensional 
architecture of eukaryotic \cite{Cremer:2006tn,Lanctot:2007kx,Misteli:2005qr} and
prokaryotic \cite{Sherratt:2003nw,Thanbichler:2005ez,Espeli:2006vw} 
chromosomes has been under active study. Yet, the fine structure at the level of 
the transcription factories and the role of chromosome architecture 
in the regulation of transcription remain to be elucidated. Several
of the important open questions are: (1) What is the mechanism
that localizes genes at their transcription factories?
(2) What is the corresponding topology of the 3-dimensional
chromosomal structure? (3) Have gene positions along DNA been selected 
during evolution so 
that they can be more easily co-localized in space during transcription?
In this article, we propose a general framework to address these questions. 

%
Let us first recall the 
two main scenarios that have been proposed for the topological organization
of chromosomes and transcription factories. In the {\it solenoid}
framework~\cite{Kepes:2003bh}, the chromosome
forms a ring, torus or solenoid, visiting the 
different foci periodically. The foci result from the bridging of distant binding sites {\it via} the binding of bivalent transcription factors. 
In support of this scenario, genes regulated by the same TFs in yeast and {\it Escherichia coli}, plus genes belonging to phylogenetically conserved gene 
pairs in {\it E. coli} have been shown to arise with some periodicity along the DNA~\cite{Kepes:2003by,Kepes:2004kb,Wright:2007rt}. A solenoidal pattern would 
then generate higher local concentrations of the TF and of its DNA binding sites, and thus might allow more efficient transcriptional 
regulation~\cite{Kepes:2003bh}; if so, this should translate into a selective advantage by analogy to the {\it lac} 
operon case~\cite{Muller-Hill:1998db}. In another framework, hereafter referred
to as the {\it rosette} scenario~\cite{Cook:2002bv}, the DNA chain first forms loops around one
single transcription factory; then, a succession of such rosettes might form a necklace of rosettes. Interactions between DNA bound proteins, 
and depletion forces due to the presence of large complexes (the transcription factories) that are surrounded by numerous small entities 
(from water molecules to proteins), have been proposed to be responsible for the formation of DNA loops~\cite{Marenduzzo:2007sr}. From a regulatory point of view, the 3-dimensional
structure of DNA has been proposed to modulate the transcription process according to the position of the genes within the loops \cite{Bartlett:2006ab}.

In both scenarios, TFs are expected to play a crucial role since, just as in the {\it lac} operon case, bivalent TFs can bridge distant sites; a multimerized form of TFs can also facilitate the bridging  \cite{MastrangeloCourey_91,ZellerGriffiths_95}; the interaction with the transcription factories, and more generally with active RNA polymerases \cite{Ptashne:1997p674}, is also expected to induce the bridging. In fact this may be so even if the binding sites are very distant
when measured along the one-dimensional DNA, in direct analogy with the numerous examples of the stabilization of DNA loops via 
the binding of bivalent \cite{Muller-Hill:1998db,Matthews:1992df,Vilar:2003mf} or multimerized \cite{MastrangeloCourey_91,ZellerGriffiths_95,Zurla:2009p584} TFs. From a theoretical point of view, a stabilizing interaction between distant binding sites
can lead to the emergence of large agglomerates of bridged sites \cite{Sumedha:2008p671}.
Unfortunately, neither computational nor theoretical work has addressed the consequences of such bridging forces on the spatial organization of these agglomerates, and hence, on the resulting chromosomal organization.

In this work, we investigate the folding properties of a single self-avoiding polymer chain along which 
specific sites interact according to a short range potential, thus mimicking an attraction mediated
by either transcription factors or large protein complexes. In such a context, we show that transcription
factories can be viewed as the result of self-organization, 
the process consisting of a micro-phase separation between
interacting and non-interacting sites.
Using Monte-Carlo simulations, we show
that a rich variety of topologies are likely to describe the spatial
co-localization of genes.
Moreover, our results strongly suggest that if genes are to co-localize 
into families according to their function or regulatory control, a 
regular pattern of gene positions along the DNA is necessary.

\vspace*{0.5cm}

All the parameters that are necessary for understanding both the modeling framework and the subsequent biological implications are listed in Table \ref{table}. A short explanatory note is provided for each parameter.

\section*{Model}
\label{sec:frame}

Statistical properties of long DNA chains in good solvents
are accurately described by worm-like chain (WLC) models
\cite{strick:2003}. 
These types of models provide a coarse-grained description of protein-coated DNA ({\it e.g.} the eukaryotic chromatin). They are simple enough to allow some analytical treatment and to be investigated numerically. They include the typical elastic behavior of DNA, which has been measured {\it in vitro} and {\it in vitro}.

More precisely, the WLC model is defined by a bending energy 
$ E_b=  K \int_{0}^{L} \left( \frac{\partial \vec t}{\partial s} \right)^2 ds$
where 
$\partial \vec t/\partial s$ is the variation of the tangent vector along
the curvilinear abscissa $s$ of the polymer, $K$ is the bending modulus, and
$L$ is the total length of the polymer. In our study,  
we further take into account the short range electrostatic repulsion of DNA (DNA is negatively charged). Due to the screening of the charges {\it in vivo}, 
it is commonplace to model this repulsion as a hard-core potential. Our framework therefore consists of a self-avoiding WLC with a hard-core radius $r_0$. 
The persistence length $l_p$  along the polymer is defined
as the distance beyond which the WLC loses most of its orientational
order -- see Fig. \ref{fig:WLC_1type}. For an infinitely thin chain ($E_b$ is then the only energy), one has $l_p = K/k_b
T$ where $T$ is the temperature in Kelvin and $k_B$ is the Boltzmann
constant. For naked DNA, $r_0 \simeq 1$ nm;
moreover typical {\it in vivo} ionic conditions lead to
$l_p \simeq 50$ nm \cite{Hagerman:1988p595}, though $l_p$ may appear larger or
smaller due to the presence of DNA bound proteins such as histone like proteins.  In the case of
eukaryotes, one
can model the 30 nm chromatin fiber by taking $r_0=15 \; {\rm nm}$; $l_p$
can vary between 50 and 250 nm, depending on the compaction level of the
chromatin. 

Within this framework, genes along the DNA are associated with specific sites on
the polymer (Fig. \ref{fig:WLC_1type}). Part of these genes will participate to the co-localization process. In this regard, several
possible scenarios have been proposed (see the introduction). Here, we
investigate the effect of thermodynamic interactions ({\it e.g.} van der
Walls or ionic) between proteins and DNA and discuss whether these
interactions can lead to a well coordinated self-organization of the
chromosome. We therefore do not consider 
proteinic complex assemblings that require energy consumption
nor possible active forces -- {\it e.g.} induced by molecular motors -- that would drive
chromosome loci to the transcription factories. Finally, the binding of
proteins on DNA is treated implicitly, that is, two chromosome loci that can
be bridged by a proteinic complex interact according to a short-range
attractive potential
%
%
$V(r)=-V_0 \times \theta(d^*+2r_0-r)$.
%
Here $\theta(x)$ is the step function that
is $1$ if $x>0$ and $0$ otherwise, $d^*$ is the interaction range, and $V_0$
is the strength of the potential.
This interaction mimics a free energy term resulting from the bridging of 
the chromosome, either by bivalent TFs such as the Lac 
repressor~\cite{Muller-Hill:1998db}, 
or by TF multimerization such as in the $\lambda$ phage \cite{Zurla:2009p584};
similarly, tethering may be mediated by the transcription
factories, or more generally by RNA polymerase/TF complexes, which occur for instance during the transcriptional activation of some bacterial genes \cite{Martin:2002p588}; values of $d^*$ therefore lie between several nanometers and several tens of nanometers. 
The free energy gain comprises that due to protein-DNA binding and, when multimerization comes into play, that 
due to protein complex formation. In any case, 
free energies ({\it i.e.}, $V_0$) are expected to be a few
kcal/mol (and thus a few $k_B T$)~\cite{Zurla:2009p584,Vilar:2003mf}.

Our coarse-graining procedure allows to tackle, within the same formalism, different mechanisms that may lead a gene to be an interacting site. For instance, our model can mimic the effect of the chromatin condensation (heterochromatin), which can prevent a site from participating to the interaction just by hiding it or making it unaccessible.
A more realistic modeling of chromosome structuration would include
heterogeneities in the interaction between the sites (different $V_0$ for
different pairs of sites) and also the explicit presence of solvent
molecules. However, our goal here is to provide a plausible general picture for
the formation of transcription factories that can be cast within a formalism as simple as possible. 

Overall, our framework consists of a self-avoiding WLC along which specific
sites are distributed sparsely and are able to interact (Fig. \ref{fig:WLC_1type}). In this context, we define $\Delta$ as the mean
distance between two successive interacting sites along the DNA.
We also define the capacity of a site as the number of other sites it can interact
with simultaneously. For the results shown here, we take for simplicity no
limit on the capacity. The maximum number of partners of a site, hereafter referred to as $\bar{n}$, will be limited only by the steric constraint: one cannot pack more than some maximum number of sites within a given distance of a point. Notice that the possibility of multiple interactions is compatible with the fact that gene regulatory regions frequently have several TF binding sites. Moreover, large protein complexes, which are likely to appear around transcription factories, should favor the simultaneous interaction of several binding sites.

\vspace*{0.5cm}


Our results can be divided into three parts. First, we show that transcription factories can be viewed as the result of a micro-structuration mechanism, which is an archetype of a self-organizing process. In particular, our calculation highlights the range of parameters for which the micro-structuration is expected. Next, we use numerical simulations to address the topological ordering of DNA around the transcription factories. Finally, we tackle the problem of forming transcription factories in the presence of different families of interacting sites, {\it i.e.} families corresponding to different regulatory properties.

\section*{Results}

\subsection*{Transcription factories as a micro-phase separation}

Before dealing with the mechanisms that are responsible for the formation of discrete foci, we quickly recall the basic phenomenology of a self-attracting and self-avoiding WLC. Within the framework of our model, this corresponds to considering a dense distribution of interacting sites along the DNA, so the interacting sites are close-by along the whole DNA. Such WLCs have been extensively studied for more than forty years \cite{de-gennes:1979ij,Grosberg:1997dc}. Depending on the values of the parameters, they mainly lead to three typical conformations, which are also known to arise for chromosomes {\it in vitro} and {\it in vivo}~\cite{Bloomfield:1997vh,Englander:2004dq}. First, in the absence of the self-attracting interaction, the WLC behaves as a self-avoiding random walk, at least on length scales larger than the persistence length; this leads to the so-called ``swollen'' state -- in the 
physics of an isolated polymer chain, some phases may arise 
only when the polymer is short; as suggested in \cite{stukan:2003tg}, we 
refer to these pseudo-phases as ``states''. Second, introduce an attractive interaction. For a sufficiently strong 
attraction, the polymer goes to one of two possible compact conformations. The densest conformation is obtained by 
having the polymer wind many times around a circle, forming in effect a kind of torus; accordingly, this has been 
called the ``toroidal'' state \cite{Grosberg:1997dc}. 
For a weaker interaction, less dense and less ordered conformations arise, the so-called "globules" where the polymer forms a
ball but otherwise seems rather random. Which of these macroscopic states -- swollen, toroidal or globule --
describes the
equilibrium state depends on the parameters, the two most important ones being the attractive force and the polymer stiffness \cite{stukan:2003tg}.

Coming back to our system consisting of a single chain with sparse interacting sites, its
peculiarity is that only {\it a few designated sites} of the chain are subject to the attraction; this means that a further organization of the chain on smaller length scales can arise as now explained.
Suppose that the interacting sites are sparsely distributed along the polymer. Starting in one of the compact states, the energy 
can be enhanced by {\it local} rearrangements, keeping the polymer compactness roughly unchanged. At a coarse-grained level, one can
focus on the local density of the interacting sites in three-dimensional space. In a random conformation, the density will be uniform. By contrast, after local rearrangements, the density will vary, leading to clumping in some areas, and voids in others. In essence, a uniform density is energetically unstable, and so the system will spontaneously structure so as to form regions of high and low densities of interacting sites (Fig.~\ref{fig:micro_trans}). This leads to 
a micro-phase separation between interacting and non-interacting sites, which is reminiscent of what is observed in block co-polymers \cite{Grosberg:1997dc}. 

In the following,  we investigate
in detail this micro-structuration, tackling the problem in two ways. First, we use a mean-field theory of polymer physics. This allows us to 
qualitatively capture the transition between the homogeneous states with a uniform distribution of interacting sites and the micro-structured states with a spatially modulated distribution of interacting sites. Next, we use Monte-Carlo simulations to both validate our analytical results and to further study the DNA conformations around the foci.

\subsection*{State diagram in biologically relevant situations}
\label{sec:micro}

Within the scope of chromosome structuration {\it via} the bridging of co-regulated genes, the macroscopic state diagram 
for not too strong attractive forces is limited to two states: the swollen state and the micro-structured globule -- see Fig. \ref{fig:phasediagprincip}. Generally speaking, the micro-structured globule tends to be favored thermodynamically over the homogeneous globule for interacting sites that are sparsely distributed along the WLC; the homogeneous density of binding sites is unstable to a modulation, at least if the capacity of sites is not too small. In this situation, the number of interacting sites lying within the foci of the micro-structured globule ($n_I$), which is kept fixed in our calculation for the sake of simplicity, determines the position of the transition between the two states. We now present the principles of the underlying calculation, emphasizing the crucial parameters that determine the balance between the states. This allows us to discuss the mechanisms responsible for the shift between the states.

\subsubsection*{Strategy} 

The best way to determine the thermodynamically favored state is to compute the free energy of each state as a function of the model parameters, which is explicitly done in section 3 in Text S1; the state with the lowest free energy is the favored one.
In the following, for the sake of simplicity, we do not tackle  the issue of the toroidal phase, considering only the micro-structured globule, the swollen state and the homogeneous globule. Ignoring their internal structure, these isotropic states look like balls. As a consequence, they can be characterized by a radius $R$ and a free energy $F=F(R)$. In the most general case, the free energy can be decomposed into four terms:
\be
F(R)=F_a(R)+F_b(R)+F_{ev}(R)+F_s(R)
\label{free}
\ee
$F_{\rm a}$ is the free energy due to the attractive potential between the interacting sites. $F_b$ is the contribution from the bending energy. 
$F_{ev}$ is the free energy cost due to the excluded volume of the polymer within an area of extension $R$, which stems 
from the repulsion of the hard-core monomers constituting the polymer.  $F_s$ is the entropy related to the number of polymer configurations that are compatible with a radius $R$ \cite{Grosberg:1997dc}.

For a given type of organization ({\it e.g.} a micro-structured globule), the free energy calculation consists in first determining the $R_*$ that minimizes the free energy. Then, $R_*$ is plugged into the free energy relation Eq. (\ref{free}), which gives the corresponding free energy of the state, that is $F^*=F(R_*)$. One must compare the free energies $F^*$ of each state. The explicit dependence on the radius $R$ of each term is calculated using a standard mean-field single chain polymer theory, also known as Flory theory  \cite{Grosberg:1997dc}, focusing on the bulk contribution to the free energies (large $R$). In the following, we skip the technical details and give the final results of the calculation, as well as its interpretation. For more details on the derivation, we refer the reader to section 3 in Text S1.

In the case of a sparse distribution of interacting sites, the position of the sites may be crucial, as we shall see in the last section. To simplify our discussion, we therefore consider, in a first stage, sites that are regularly spaced by $\Delta$ along the DNA.

\subsubsection*{Homogeneous states}

The calculation shows that the balance between the homogeneous globule and the swollen states rests on the value of 
the single parameter (see sections 3.3 and 3.4 in Text S1):
\ben
\frac{\bar{n}V_0}{k_B T}  \times \frac{\Delta}{l_p} \times \left(\frac{r_0}{\Delta}\right)^{3} 
\label{swo}
\een
Accordingly, three mechanism can be responsible for the greater stability of one state compared to the other. First, there is the competition between the attractive potential coming from the interacting sites on the one hand, and the destabilizing thermal energy coming from the solvent on the other. This corresponds to the term $\bar{n}V_0/k_B T $; notice that the effective free energy of attraction per site is proportional to the maximum number of partners of a site ($\bar n$). Second, $\Delta/l_p$ reflects the difficulty for rigid polymers (large $l_p$) to bridge interacting sites that are close by along the polymer. Finally, the ability of the polymer to form contacts between interacting sites crucially depends on the number of these sites. This is reflected by the term $ r_0 / \Delta$, which corresponds to the linear density of the interacting sites along the polymer. 

Thus, the swollen state is more stable whenever the above parameter is small compared to 1, that is,  at high temperature, for rigid polymers, and when few interacting sites are present along the polymer. In the opposite case, \emph{i.e.},  at sufficiently low temperatures, for sufficiently flexible polymers, and for a sufficient number of interacting sites, the homogeneous globule becomes more stable. 

\subsubsection*{The micro-structured globule} 

As we shall see below, in biological situations in which transcriptionally regulated genes are involved, the swollen state is always more stable than the homogeneous globule. In these conditions, the micro-structured globule is the most stable state if and only if it is more stable than the swollen state. To tackle this point, for the sake of simplicity, we suppose: i) that the foci are composed of $n_I$ interacting sites where $n_I$ is uniform across the whole globule, and ii) that two nearest-neighbor foci are separated by a distance that is also uniform across the whole
globule. These hypotheses are mean-field-like since they neglect spatial variations of certain characteristics of the polymer. For $\Delta/l_p$ not too small, which is appropriate for gene co-localization (see below), one can show that stability depends on the value of the single parameter (see section 3.5 in Text S1): 
\be
n_I^{2/5} \left( \frac{\Delta}{l_p} \right)^{1/5} \frac{k_B T}{\bar{n} V_0}
\label{het}
\ee
For low (respectively high) values of this parameter, {\it i.e.}, when $n_I^{2/5} \left( \Delta/ l_p \right)^{1/5} (k_B T/\bar{n} V_0)$ 
is much smaller (respectively much larger) than $1$, the micro-structured globule is more (respectively less) stable. 

Three ingredients are therefore crucial for the stability of the micro-structured globule. First, big foci (large $n_I$'s) tend to be less stable than small foci, although a rigorous calculation would require $n_I$ not to be fixed {\it a priori}: 
nothing prevents foci from splitting if this lowers their free energy. 
Notice that the number of sites per foci is expected to be limited from above by the hard-core properties of the polymer ($r_0$), and by the properties of the interaction as well ($d^*$). This can be checked by numerical simulations.

Next, some amount of rigidity seems to be necessary in order to stabilize the micro-structured globule since small values of $\Delta/l_p$ tend to lower the value of the above parameter. This may appear counter-intuitive with respect to what has been stated in the previous section, namely, that rigid polymers tend to favor swollen states. This last statement is true but the results presented in this section are valid only when $\Delta/l_p$ is not too small, \emph{i.e.}, when $\Delta/l_p \gtrsim 1$. In this limit, which is the one of biological relevance to our problem, the more rigid the polymer,  the lower the excluded volume coming from the non-attracting parts of the polymer. Indeed, little space is available for the polymer to fluctuate in between the foci. Hence, a rigid polymer would tend to diminish the fluctuations so that the hard-core repulsions between the monomers would diminish (with an increase of the distance between the foci). Overall, this would
  tend to stabilize the micro-structured globule. Nevertheless, very large values of the rigidity, {\it i.e.} $\Delta/l_p \ll 1$, would eventually destabilize the micro-structured globule to give way to the swollen state. In any case, due to the low value of the exponent $1/5$ in relation (\ref{het}), the effect of varying $\Delta/l_p$ on the state diagram is rather modest, at least in the biological situations we are interested in -- see Fig. \ref{fig:phasediagprincip}. Finally, the above parameter shows that strong attracting interactions ($k_B T/\bar{n} V_0 \ll 1$) naturally tend to favor the micro-structured globule.

\subsection*{Application: gene co-localization and transcriptional regulation}  

Our WLC offers a single framework to discuss the formation of transcription factories both in bacteria and in eukaryotes. Within the context of transcriptional regulation, genes participating to the same transcription factories are believed to participate to specific cellular functions. $\Delta$ can therefore be evaluated as the typical distance between two consecutive genes that are co-regulated by the same TF or that are known to participate to the same function. As a consequence, $\Delta$ is expected to be larger than the distance separating two genes, {\it e.g.} $\sim$ 1 kbps  in bacteria ({\it i.e.},
$\Delta \geq$ 300 nm), and $\sim$ 100 kbps in mammals ($\Delta \geq$ 600 nm)  -- we have
used 150 bps/nm for the chromatin fiber \cite{Langowski:2006p648}. This leads to factors $(\Delta/r_0)^{-3} \ll 1$ in relation (\ref{swo}). Hence, within the scope of our model, for biologically relevant values of $\bar{n}$ and $V_0$, the homogeneous globule state
(with a uniform distribution of actively transcribed genes) is thermodynamically unlikely both in bacteria and eukaryotes.

As far as the micro-structured globule with discrete foci is concerned, in eukaryotes one can approximate $n_I$ as the
typical number of active RNA polymerases within one transcription factory, {\it i.e.}, $n_I \sim 30$ \cite{Jackson:1998yw}. By considering $l_p=250$ nm and $\Delta=1$ Mbps (6 mm), which corresponds to the mean distance between two consecutive genes regulated by a TF in the human genome \cite{Vaquerizas:2009p562}, one finds $n_I^{2/5} \left( \Delta/ l_p \right)^{1/5} (k_B T/\bar{n} V_0) \simeq 7 {k_B T}/{\bar{n} V_0}$. 
The regulatory regions of eukaryotic genes often have several TF binding
sites of the same type, which can be interpreted as $\bar{n} \gtrsim
2$. Hence, a bridging induced by TFs (with binding energies of several $k_B T$'s per
TF), or induced by a proteinic complex involving TFs, is sufficient to induce the formation of transcription factories
according to the above micro-phase separation ($n_I^{2/5} \left( \Delta/ l_p \right)^{1/5} (k_B T/\bar{n} V_0) < 1$). Moreover, given the parameters of chromatin, the values of $\bar{n}$ and $V_0$ lie in a range that allows to switch between a state with discrete foci and the swollen state -- see Fig. \ref{fig:phasediagprincip}. This suggests that
the micro-phase structuration is also a possible mechanism for fine tuning the global genetic regulation of a cell.

In bacteria, an interesting case concerns the formation of the putative transcription factories during
the transcription of rRNA operons \cite{Cabrera:2003fu}. In this situation,
7 operons scattered along 2 Mbps have to be co-localized. $l_p \sim 50$ nm
then leads to $n_I^{2/5} \left( \Delta/ l_p \right)^{1/5} (k_B T/\bar{n} V_0) \simeq 10 {k_B T}/{\bar{n} V_0}$. In the same
way, both co-regulated genes  \cite{Kepes:2003by} and genes that are thought
to be functionally related \cite{Wright:2007rt} have been shown to be
periodically spaced according to a $\Delta$ $\sim$ 100 kbps
period. Supposing these genes are co-localized by groups of at least ten,
this leads again to $n_I^{2/5} \left( \Delta/ l_p \right)^{1/5} (k_B T/\bar{n} V_0) \simeq  10 {k_B T}/{\bar{n} V_0}$. Hence, in
bacteria, if one considers one single binding site per gene, large binding
energies are required for the formation of transcription factories. However,
this should be balanced by the overall negative supercoiling of bacterial DNA
which is beyond the scope of our model. Together with the action of nucleoid-associated
proteins ({\it e.g.} histone like proteins such as Fis, H-NS or HU), this effect would tend to condense the
chromosome and hence to dampen consequences of thermal fluctuations.

\subsection*{DNA organization of the micro-structured globule}
\label{sec:different}

Numerical simulations of polymer models are useful to investigate the principles of chromosome organization within space \cite{Marenduzzo:2007sr,Rosa:2008p511,Nicodemi:2009p638}.
In this respect, simulations of our self-avoiding WLC (see Methods for details) confirm that gene foci arise
for  persistence lengths, binding free energies and inter-gene distances
that are typical of bacteria and eukaryotes (see
Fig. \ref{fig:traveling} for two such examples). Simulations are also
useful to see how the foci organize in 3-dimensional space. Indeed, \emph{a priori}, foci may
form regular lattices, random lattices, or they may wander with time. In this respect, 
our results suggest a rich variety of equilibrium
conformations that depend on the parameters of the system. However, from a
computational point of view, we are not able to investigate the {\it thermodynamic} state
diagram when the DNA chain becomes relatively large
because the different metastable states last the whole time window of the
simulation once they are formed; in particular, we do not see switches
between the states as would arise in a situation of
co-existence. Thus we are limited to considering the most likely structures that form when
starting from a random coil (swollen) configuration as we progressively
increase the value $V_0$ from an initial zero value. Note that from a biological
point of view, such metastable states may
be just as relevant as the true equilibrium states.

The resulting structures can be divided into three main groups, as we now describe.

\subsubsection*{The micro-structured solenoids}

Begin with a toroidal conformation of a self-attracting WLC and try
to maximize the number of sites in interaction when $\Delta/l_p$ increases. To
do that, one can take the sites and push/slide them so that they co-localize
in sections of the torus, {\it i.e.}, agglomerate in foci along planes that cut
the torus through its small section (Fig. \ref{fig:microphases}a). One can
obtain ring-like structures or open linear structures that are topologically
equivalent. To differentiate these structures from the uniform toroidal
conformations, we refer to them as {\it solenoidal}. This type of
organization has been advocated by K\'ep\`es in 2003 to justify the tendency for
genes that are co-regulated by the same TFs to be periodically positioned
along the DNA \cite{Kepes:2003bh}, and by Wright {\it et al.} in 2007
to explain periodic trends in the position of phylogenetically conserved
gene pairs in bacteria \cite{Wright:2007rt}. In the limit of extremely
compact DNA, the number of planes is determined by the periodicity parameter
$\Delta/l_p$ and by the size of the torus. As can be seen in 
Fig. \ref{fig:solenoid_rosette} from the numerical simulation of the WLC, such rings of 
gene foci and topologically equivalent open conformations arise for some parameters 
of the WLC that are relevant for describing naked DNA.

\subsubsection*{The rosette structure}

If $\Delta/l_p$ is increased, one can reduce the number of foci by putting
more binding sites in each. At some point, there will remain a single focus
if it has the capacity to hold all the binding sites. In this situation, the
polymer performs round trips about one single focus as shown in
Figure S2. In the case of several foci, it may be that 
successive interacting sites on the polymer belong to the same focus before
going on to another one (Fig. \ref{fig:microphases}c). This is the kind of structure advocated by Cook
et al.~\cite{Cook:2002bv} for DNA organization around transcription
factories, the loops of DNA being tethered to one focus. We shall call it a
``necklace of rosettes'' because each focus corresponds to a rosette. Do
such structures arise in our simulations of the WLC? For most of the
parameter values we studied, we have never observed more than 2 successive
rosettes (see Fig.~\ref{fig:solenoid_rosette} for such a situation
when parameters are set to correspond to the eukaryotic case). 
However, more rosettes may arise when several types of interaction are present, as we shall see in the next section. 

\subsubsection*{The traveling chain  structure}

The previous solenoidal and rosette structures 
have a significant entropic cost; if energetic effects cannot compensate this, one expects
these two structures to be destabilized.
As reported in Figs.~\ref{fig:traveling} and \ref{fig:solenoid_rosette}, 
our numerical results show that for long polymers,
bundles of chain segments free of interacting sites may form a spatial network 
while the interacting sites are concentrated within
the nodes (see Video S1 showing the
three-dimensional realization). Within this network, when going from one
binding site to the next one along the DNA chain, one typically moves to a
different focus. Accordingly, we call this structure ``the traveling chain''
structure.

\subsubsection*{Topological state diagram}

A schematic view of the resulting state diagram is depicted in Fig. \ref{fig:diagramMC}.  
Interestingly, the three topological orderings (rosette, solenoid, traveling chain)
can also be distinguished mathematically by a novel order parameter. Its construction is
based on following the successive DNA binding sites which belong to various foci (see section 4.1 in Text S1 for the mathematical formulation); the
successive steps 
can be thought of as a random walk, leading to 3 kinds of behaviors.
For the necklace of rosettes, the walk visits the same focus multiple times
but then goes away "for ever": the random walk is "transient". The other two cases
correspond to "recurrent" random walks. 
In the case of the toroidal ordering, the walk does not visit the
same focus twice in a row but comes back to the same focus after a (large) number
of steps (the recurrence property). The
traveling chain also gives a recurrent walk, but in contrast to the 
toroidal case, has a finite probability of revisiting
the same focus within a few steps. 
Notice also that the interacting sites do not
necessarily co-localize into spherical foci. Depending on the parameters, they can also organize themselves according to one-dimensional shapes (Figure S3).

Finally, recall that to simplify our study, we have used
interacting sites that were periodically spaced along the DNA. Our results
are robust to deviations from this case: small amounts of disorder do not change 
the possible states -- see section 4.1 in Text S1. In the case of 
a fully random distribution of the gene positions, 
for small contour length $L$ we observe rosette structures (with still one
or two foci only) whereas large $L$ seems to favor the formation of spatial
networks of foci. 

\subsection*{Generalization to multiple kinds of binding sites}
\label{sec:transcription}

Recent experiments in monkey Cos7 cells have shown that different
transcription factories recruit different genes depending on their
promoter type \cite{Xu:2008dk}. In the same spirit, one may hypothesize
that genes regulated by the same TFs preferentially co-localize
in space \cite{Kepes:2003bh,Osborne:2004pi,Simonis:2006dk}. This would explain 
for instance, in yeast, the tendency of co-regulated genes to be clustered along the chromosomes \cite{Wagner:1999p785,Cohen:2000p786,Kepes:2003bh}.
A somewhat analogous issue arises in bacteria: one often finds that a 
TF coding gene, the binding site of that TF, and the corresponding 
regulated gene(s) are all close-by along the DNA (see \cite{Kepes:2004kb} and references therein).
This is thought to optimize the
three-dimensional targeting process of the TF toward its binding site
because, in bacteria, protein translation occurs close to the coding
gene. Accordingly, space co-localization of distant binding sites 
 for each TF type may very well occur since it is a natural way to make
three-dimensional targeting and assembly of complexes
more efficient. The investigation of gene
positions in {\it E. coli} and yeast suggests that in these organisms a near
periodic arrangement on the DNA of co-regulated genes may be at the base
of a good 3-dimensional spatial 
co-localization \cite{Kepes:2003by,Kepes:2004kb}. 
However, there are hundreds of TF types both in bacteria and yeast,
and thousands in higher eukaryotes so that the
satisfaction of all the separate co-localization constraints may be a hard problem for the organism to solve.

We have used our framework to numerically model the
spatial co-localization process when $N_t$ different types of TFs regulate a
large number of genes. Specifically, we have $N_t$ types of binding sites,
where two binding sites interact only if they are of the same type. The way
these sites (and their types) are positioned along the chain can affect the
way the different foci form. 
We have therefore compared the co-localization process using four kinds of
positioning of these binding sites, namely: i) sites ordered -- and thus 
clustered -- according to their types ii) randomly distributed sites
and types; iii) periodically distributed sites and types; iv) sites that are
spaced according to random multiples of $l_p$, hereafter referred to
as {\it random periodic}: there is approximate 
periodicity in the site positions while the site types are taken to be
completely random (see Figure S4 for an illustrative
explanation). Situation $i$ corresponds to the one-dimensional clustering of nearby binding sites whereas situations $ii$ to $iv$ correspond to the interaction of binding sites that can be distant from each other. In particular, situation $iv$ is useful
to determine whether regularity in the 
site types is necessary for co-localization, even if there is some
regularity in the site positions along the DNA.
In this context, the mean distance $\lambda$ (measured along the
chain) between two consecutive sites regardless of their type is a useful
additional parameter to characterize the one-dimensional site properties
along the DNA. To simplify our study, we take a number of interacting sites
that is roughly the same for each site type so that $\lambda \simeq \Delta /
N_t$, $\Delta$ being the mean distance between two sites of the same type.

\subsubsection*{One-dimensional clustering {\it vs.} periodic spacing induced topologies} 

The topological organization of DNA in response to the activation of transcription can dramatically depend on the
organization of genes along the chromosome. Fig. \ref{fig:rosettevssol} reports typical chromosome configurations, both  for chromatin and naked DNA, that are obtained when the sites are ordered according to their types (case $i$) and when they are periodically spaced (case $iii$). Both genomic organizations lead to a rapid formation of homogeneous transcription factories in space. 
Periodic spacing induces the formation of solenoidal configurations while one-dimensional clustering induces the formation of necklaces of rosettes.

\subsubsection*{Periodic site positions favor co-localization of distant sites}
Our results show that some periodicity in the site positions allows for an
efficient spatial co-localization of distant sites and also that a too disordered
positioning hinders the formation of foci. As shall be explained now, this can be seen in both the dynamic and static aspects of the folding transition of our polymer model.

First of all, the folding transition from an unstructured state to a
steady state takes a longer time in the
random case than in the periodic cases. This is illustrated in Fig.~\ref{fig:folding} using single
trajectories. In general, the larger $N_t$,
the larger this effect. Moreover, as expected, the folding times tend to
become similar when $N_t$ decreases at fixed $\lambda$ or when 
$\lambda$ increases at fixed $N_t$.

Second, the fraction $\rho(s)$ of binding sites belonging to a focus of size
$s$ in a steady state depends on the organization of the
interacting sites along the DNA (Fig. \ref{fig:folding}). For pure periodic positions, the polymer
forms either a solenoidal structure or a well organized network of foci for
long chains which succeed in clustering all the binding sites. In this
situation, $\rho(s)$ has a single peak at large sizes $s$. This pure
periodic case can be viewed as an ideal context for forming specialized
factories. Interestingly, our results further suggest that partial
periodicity in the position of the sites ({\it i.e.}, case $iv$ with an imperfect
periodic organization) is sufficient to have an efficient spatial
co-localization mechanism for which all foci have more or less the same
size, {\it i.e.}, $\rho(s)$ exhibits mainly a single peak as shown in
Fig.~\ref{fig:folding}. In contrast to the pure periodic case, the spatial
structure is not a ring-like structure although there is some circularity in
the structure (Fig. \ref{fig:orga}). When positions are 
drawn randomly (case $ii$), $\rho(s)$ becomes bimodal for large
values of $N_t$ and fixed values of either $\lambda$ or $\Delta$. 
In particular, a peak at $s=1$ appears,
the other main peak corresponding to large values of $s$. Hence, a finite
fraction of the sites remains isolated in space, {\it i.e.}, many sites do not belong to
a so-called transcription factory, even though large clusters are
formed (Figure S5). Moreover, by running different trajectories from different random
initial configurations, we have observed that in the random case the
steady states differ from run to run, {\it i.e.}, the way the sites 
cluster varies, although the parameters of the polymer model are kept
fixed (data not shown). 
Overall, the situation is reminiscent of a thermodynamic glass transition where the equilibrium free energy is
dominated by multiple thermodynamic states that are separated by high energy
barriers. In such situations, frustration, which is due here to the presence of
binding sites along the polymer that are incompatible with a ring-like
structure, is a crucial feature for constraining the thermodynamic state.

Third and lastly, it is interesting to compare the spatial conformations of the
structured state for the random and the periodic positioning of the
interacting sites. As can be seen from Fig. \ref{fig:orga}, if
positions are randomly drawn, the clusters tend to form close to each other
in space whereas in the pure periodic case, the clusters are well separated
and are periodically spaced along a torus. Moreover, in the random periodic
case, the clusters are also well separated although no specific ring
structure is formed. Overall, these results show that some regularity in the
positions of distant interacting sites is needed to have 
well separated foci in space,
which presumably is a pre-requisite for a good operation 
of transcription factories.

\section*{Discussion}
\label{sec:discussion}

Within a fairly general framework, we investigated the topological organization
of a model chromosome. Using an effective attractive potential between
selected genes on a DNA chain, we found that these could organize into
discrete foci, with the DNA visiting the foci in several topologically
distinguishable ways. The foci are composed of genes that 
can be far away from each other along the DNA, which is supported by 
the recent observation of numerous Mbps-range DNA 
loops~\cite{Osborne:2004pi,Simonis:2006dk}. 

Of course, {\it in vivo}, numerous obstacles might prevent chromosomes from 
achieving the conformations we predict: supercoiling, chromatin
remodeling and confinement introduce other interactions that may dominate for some 
parameter values. 
Another point is that we have focused on equilibrium conformations
whereas in reality cellular processes  
operate away from equilibrium. However, a pure equilibrium approach is useful because it shows the natural organizational trend of the system. 

Several conclusions transpire from our framework.
First, in bacteria and eukaryotes, the formation of transcription factories
may be related to a self-organizing process akin to the folding
transition of single polymer chains.
The underlying thermodynamic mechanism is a spatial
micro-phase separation driven by regions of DNA where genes are 
subject to similar transcriptional regulation. In effect, due to the 
very nature of the self-avoiding DNA chain, 
all genes cannot cluster together to enhance transcription rates; instead,
discrete foci must form in space. Our results therefore confirm that self-organization may play a crucial role in the structuring of chromosomes \cite{Misteli:2005qr,Kosak:2007p607}.

Interestingly, the interaction strength needed
between distant sites along the DNA in order to induce the micro-structuration is compatible with the binding of TFs to DNA. The bridging can be achieved {\it via} a bivalent TF, or more generally through the formation of large protein complexes, {\it e.g.} by tethering the DNA-bound TF to ongoing transcription factories. This corroborates 
TFs as possible entities for mediating the effective attractive
potential; our model therefore 
predicts a 3D co-localization of co-regulated genes. In eukaryotes, this can be tested by a combination 
of 3D fluorescence in situ hybridizations (FISH) and chromosome conformation
capture techniques~\cite{Dostie:2006p561} as exemplified in \cite{Osborne:2004pi,Simonis:2006dk,Dostie:2006p561,Jhunjhunwala:2008p599}. 
In bacteria, this can be tested by using the site-specific recombination system of the bacteriophage $\lambda$ \cite{Espeli:2006vw}.
Furthermore, as illustrated by Eq. (\ref{het}),
the number of co-regulated genes that can be co-localized within the same focus
depends both on the number of TF binding sites per gene and on the binding
energies. This leads to the prediction
that the presence of aptamers which can compete with TFs for binding to
cognate DNA sites will lead to smaller transcription factories or
even none at all.

Second, using numerical simulations of our model, we have shown that the 
topology of the DNA conformations fall into several classes according to the
way the foci are visited, and that two of these classes 
had been previously hypothesized on the basis of biological evidence. For
instance, starting from a toroidal organization of DNA which has been
observed in some organisms \cite{Englander:2004dq},
if the interacting sites that stabilize this structure become less dense, 
there should be a micro-phase separation whereby distinct foci appear 
along the ring, which fits the solenoidal model 
proposed in~\cite{Kepes:2003bh}. 
As interacting site density decreases further, rosettes may
form as proposed in ~\cite{Cook:2002bv}. Or the DNA may successively visit the different
foci in a random fashion, corresponding to our ``traveling chain''
topology. 

Third, which topological ordering arises generally depends
on the way the binding sites are positioned along the one-dimensional
DNA. We find that some periodic regularities and some clustering in the positioning of co-regulated genes, as observed respectively in ~\cite{Kepes:2003by,Kepes:2004kb} and in {\it e.g.} \cite{Wagner:1999p785,Cohen:2000p786}, strongly favor the formation of well-separated foci with a homogeneous size and content, and disfavor the presence of genes outside of the foci.
To this end, we considered the possibility of
having multiple types of protein binding sites, thought to be
associated with different transcription factor families or gene functions.
We found that having periodically-positioned targets of multiple TFs favored the solenoidal topology whereas the necklace of rosettes topology
was favored if groups of genes were one-dimensionally clustered
along the DNA (Fig. \ref{fig:rosettevssol}).



\section*{Methods}

\subsection*{Numerical implementation of the WLC model}

Numerical simulations of the continuous self-avoiding WLC model are based on an off-lattice semi-flexible polymer composed of $N$ jointed
cylinders of radius $r_0$ and length $a_0$ (Figure S1). The cylinders are impenetrable (hard-core interactions) and two consecutive cylinders $i$, $i+1$ that form a bending angle $\delta_i$ contribute a bending energy $E_b^i=\frac{K}{2a_0}
\delta_i^2$ to the total energy $E$. The solvent is implicit, it is not treated explicitly.

Interacting sites are taken to be located at the joints between two consecutive cylinders; a joint can contain or not an interacting site. 
They interact {\it via} a uniform short range square potential of depth $-V_0$ and interaction range $d^*$ (Figure S1). Thus, if two non-consecutive interacting sites $I$ and $J$ can interact, they contribute an energy $-V_0$ if the distance $r_{IJ}$ between them is less than $d^*+2r_0$. As a result, the total energy of the system reads:
\be
E=\frac{K}{2a_0} \sum_{i=0}^{N-1} \delta_i^2-V_0\sum_{\langle I,J \rangle} \theta(d^*+2r_0-r_{IJ})
\ee
where $\langle I,J \rangle$ means that the non-consecutive interacting sites $I$ and $J$ are able to interact. $\theta(x)$ is the step function that is equal
to $1$ if $x>0$ and is $0$ otherwise. $K$ is the bending modulus of the polymer;
it depends on the type of polymer (DNA or chromatin) that is described. To
have results that are insensitive to the discrete nature of the polymer
representation, one should use a segment length that is a small fraction of the persistence length; in all our simulations, we take this factor to be one fifth. Note also that it is best to work {\it off-lattice} as lattice anisotropy is known to induce geometrical artifacts that could spoil the interpretation of the results. 

The persistence length and the radius of our 
polymer representation of DNA depend on the type of 
organism to be modeled. In the limit of an infinitely thin 
polymer ($r_0 \to 0$), the persistence length $l_p$ is
related to the bending modulus via $l_p = K/k_B T$. 
In the case of the self-avoiding polymers presented in this work, 
this relation holds well (data not shown) so that the bending energy is enough to define $l_p$.

\subsection*{Monte Carlo simulation}

To sample the state space of our polymer model, we use standard Monte Carlo
procedures with the Metropolis accept/rejection rule, which guarantees reaching
thermodynamic equilibrium if ergodicity is not broken. The Monte Carlo
method consists in 1) picking at random two joints (a joint being the point
where two consecutive cylinders coincide), and 2) applying a 3-dimensional
rotation around the axis that passes through the two joints according to a
random angle in $[-\lambda_m;\lambda_m]$. Here we take a relatively small
value, $\lambda_m = \pi/10$ (at larger values the acceptance rate goes
down).

\subsection*{Polymer time scales and steady states}

The largest timescale for the conformational relaxation of a single coiled
polymer scales as $L^2$  \cite{de-gennes:1979ij}. From a numerical point of
view, this results in a relaxation times that scales as $N^2$ for local
microscopic evolution rules, \emph{i.e.}, where only a finite number of cylinders
are updated at each time step. In our Monte Carlo simulation, time is
counted in number of sweeps, one sweep consisting of $N$ attempts to rotate part of 
the chain; also $O(N)$ monomers are updated during one single
rotation. This leads to a relaxation time that scales as $N$. Nevertheless, we
still need roughly $N^2$ computer operations to thermalize the system
in the regime of interest where interaction effects are dominant. This
prevents the current method from scaling up to very long chromosomes,
although we can deal with interesting systems. We present simulations with up to $N=1000$ cylinders; 
this corresponds to $\sim$ 50 kbps in the case of naked DNA and $\sim$ 5 Mbps in the case of the chromatin fiber. In this case, we are not able to sample the equilibrium space of the condensed polymer because the different metastable states are very stable.

In situations of slow temporal evolution, defining a steady state may be a tricky operation. For the parameters we used, our results suggest to consider as steady a state that lasts for more than $2 \times 10^6$ sweeps. For random positions of the interacting sites, the folding time can exceed $2 \times 10^6$ sweeps. Notice then that $2 \times 10^6$ sweeps correspond to $2 \times 10^9$ Monte-Carlo steps for $N=1000$ (the largest size we report here).

\section*{Acknowledgments}
We thank G. Bianconi, M. Marsili and M. Weigt for helpful comments. This work was supported by the Sixth European Research 
Framework (project number 034952, GENNETEC project), PRES UniverSud Paris, CNRS and Genopole.

\bibliography{bib}

\begin{thebibliography}{10}
\providecommand{\url}[1]{\texttt{#1}}
\providecommand{\urlprefix}{URL }
\expandafter\ifx\csname urlstyle\endcsname\relax
  \providecommand{\doi}[1]{doi:\discretionary{}{}{}#1}\else
  \providecommand{\doi}{doi:\discretionary{}{}{}\begingroup
  \urlstyle{rm}\Url}\fi
\providecommand{\bibAnnoteFile}[1]{%
  \IfFileExists{#1}{\begin{quotation}\noindent\textsc{Key:} #1\\
  \textsc{Annotation:}\ \input{#1}\end{quotation}}{}}
\providecommand{\bibAnnote}[2]{%
  \begin{quotation}\noindent\textsc{Key:} #1\\
  \textsc{Annotation:}\ #2\end{quotation}}
\providecommand{\eprint}[2][]{\url{#2}}

\bibitem{Cremer:2006tn}
Cremer T, Cremer M, Dietzel S, Muller S, Solovei I, et~al. (2006) Chromosome
  territories -- a functional nuclear landscape.
\newblock Curr Opin Cell Biol 18: 307--316.
\bibAnnoteFile{Cremer:2006tn}

\bibitem{Sexton:2007yt}
Sexton T, Schober H, Fraser P, Gasser S (2007) Gene regulation through nuclear
  organization.
\newblock Nat Struct Mol Biol 14: 1049--1055.
\bibAnnoteFile{Sexton:2007yt}

\bibitem{Jackson:1993ye}
Jackson DA, Hassan AB, Errington RJ, Cook PR (1993) Visualization of focal
  sites of transcription within human nuclei.
\newblock EMBO J 12: 1059--1065.
\bibAnnoteFile{Jackson:1993ye}

\bibitem{Wansink:1993gf}
Wansink DG, Schul W, van~der Kraan I, van Steensel B, van Driel R, et~al.
  (1993) {Fluorescent labeling of nascent RNA reveals transcription by RNA
  polymerase II in domains scattered throughout the nucleus.}
\newblock J Cell Biol 122: 283--293.
\bibAnnoteFile{Wansink:1993gf}

\bibitem{Cabrera:2003fu}
Cabrera JE, Jin DJ (2003) {The distribution of RNA polymerase in {\it
  Escherichia coli} is dynamic and sensitive to environmental cues.}
\newblock Mol Microbiol 50: 1493--1505.
\bibAnnoteFile{Cabrera:2003fu}

\bibitem{Osborne:2004pi}
Osborne CS, Chakalova L, Brown KE, Carter D, Horton A, et~al. (2004) Active
  genes dynamically colocalize to shared sites of ongoing transcription.
\newblock Nat Genet 36: 1065--1071.
\bibAnnoteFile{Osborne:2004pi}

\bibitem{Simonis:2006dk}
Simonis M, Klous P, Splinter E, Moshkin Y, Willemsen R, et~al. (2006) {Nuclear
  organization of active and inactive chromatin domains uncovered by chromosome
  conformation capture-on-chip (4C).}
\newblock Nat Genet 38: 1348--1354.
\bibAnnoteFile{Simonis:2006dk}

\bibitem{Muller-Hill:1998db}
{M\"{u}ller-Hill, B} (1998) The function of auxiliary operators.
\newblock Mol Microbiol 29: 13--18.
\bibAnnoteFile{Muller-Hill:1998db}

\bibitem{Lanctot:2007kx}
{Lanct\^ot, C and Cheutin, T and Cremer, M and Cavalli, G and Cremer, T} (2007)
  Dynamic genome architecture in the nuclear space: regulation of gene
  expression in three dimensions.
\newblock Nat Rev Genet 8: 104--115.
\bibAnnoteFile{Lanctot:2007kx}

\bibitem{Misteli:2005qr}
Misteli T (2005) Concepts in nuclear architecture.
\newblock Bioessays 27: 477--487.
\bibAnnoteFile{Misteli:2005qr}

\bibitem{Sherratt:2003nw}
Sherratt DJ (2003) Bacterial chromosome dynamics.
\newblock Science 301: 780--785.
\bibAnnoteFile{Sherratt:2003nw}

\bibitem{Thanbichler:2005ez}
Thanbichler M, Wang SC, Shapiro L (2005) The bacterial nucleoid: a highly
  organized and dynamic structure.
\newblock J Cell Biochem 96: 506--521.
\bibAnnoteFile{Thanbichler:2005ez}

\bibitem{Espeli:2006vw}
{\'Esp\'eli, O and Boccard, F} (2006) {Organization of the {\it Escherichia
  coli} chromosome into macrodomains and its possible functional implications.}
\newblock J Struct Biol 156: 304--310.
\bibAnnoteFile{Espeli:2006vw}

\bibitem{Kepes:2003bh}
{K\'ep\`es, F and Vaillant, C} (2003) Transcription-based solenoidal model of
  chromosomes.
\newblock Complexus 1: 171--180.
\bibAnnoteFile{Kepes:2003bh}

\bibitem{Kepes:2003by}
{K\'ep\`es, F} (2003) Periodic epi-organization of the yeast genome revealed by
  the distribution of promoter sites.
\newblock J Mol Biol 329: 859--865.
\bibAnnoteFile{Kepes:2003by}

\bibitem{Kepes:2004kb}
{K\'ep\`es, F} (2004) {Periodic transcriptional organization of the {\it E.
  coli} genome}.
\newblock J Mol Biol 340: 957--964.
\bibAnnoteFile{Kepes:2004kb}

\bibitem{Wright:2007rt}
Wright MA, Kharchenko P, Church GM, Segre D (2007) Chromosomal periodicity of
  evolutionarily conserved gene pairs.
\newblock Proc Natl Acad Sci U S A 104: 10559--10564.
\bibAnnoteFile{Wright:2007rt}

\bibitem{Cook:2002bv}
Cook PR (2002) Predicting three-dimensional genome structure from
  transcriptional activity.
\newblock Nat Genet 32: 347--352.
\bibAnnoteFile{Cook:2002bv}

\bibitem{Marenduzzo:2007sr}
Marenduzzo D, Faro-Trindade I, Cook PR (2007) What are the molecular ties that
  maintain genomic loops?
\newblock Trends Genet 23: 126--133.
\bibAnnoteFile{Marenduzzo:2007sr}

\bibitem{Bartlett:2006ab}
Bartlett J, Blagojevic J, Carter D, Eskiw C, Fromaget M, et~al. (2006)
  Specialized transcription factories.
\newblock Biochem Soc Symp : 67--75.
\bibAnnoteFile{Bartlett:2006ab}

\bibitem{MastrangeloCourey_91}
Mastrangelo IA, Courey AJ, Wall JS, Jackson SP, Hough PV (1991) {DNA looping
  and Sp1 multimer links: a mechanism for transcriptional synergism and
  enhancement}.
\newblock Proc Natl Acad Sci U S A 88: 5670-5674.
\bibAnnoteFile{MastrangeloCourey_91}

\bibitem{ZellerGriffiths_95}
Zeller RW, Griffith JD, Moore JG, Kirchhamer CV, Britten RJ, et~al. (1995) {A
  multimerizing transcription factor of sea urchin embryos capable of looping
  DNA}.
\newblock Proc Natl Acad Sci U S A 92: 2989--2993.
\bibAnnoteFile{ZellerGriffiths_95}

\bibitem{Ptashne:1997p674}
Ptashne M, Gann A (1997) Transcriptional activation by recruitment.
\newblock Nature 386: 569--77.
\bibAnnoteFile{Ptashne:1997p674}

\bibitem{Matthews:1992df}
Matthews KS (1992) {DNA looping.}
\newblock Microbiol Rev 56: 123--136.
\bibAnnoteFile{Matthews:1992df}

\bibitem{Vilar:2003mf}
Vilar JMG, Leibler S (2003) {DNA looping and physical constraints on
  transcription regulation.}
\newblock J Mol Biol 331: 981--989.
\bibAnnoteFile{Vilar:2003mf}

\bibitem{Zurla:2009p584}
Zurla C, Manzo C, Dunlap D, Lewis D, Adhya S, et~al. (2009) {Direct
  demonstration and quantification of long-range DNA looping by the $\lambda$
  bacteriophage repressor}.
\newblock Nucleic Acids Res 37: 2789-2795.
\bibAnnoteFile{Zurla:2009p584}

\bibitem{Sumedha:2008p671}
Sumedha, Weigt M (2008) {A thermodynamic model for the agglomeration of
  DNA-looping proteins}.
\newblock J Stat Mech 11: 005.
\bibAnnoteFile{Sumedha:2008p671}

\bibitem{strick:2003}
Strick TR, Dessinges MN, Charvin G, Dekker NK, Allemand JF, et~al. (2003)
  Stretching of macromolecules and proteins.
\newblock Rep Prog Phys 66: 1-45.
\bibAnnoteFile{strick:2003}

\bibitem{Hagerman:1988p595}
Hagerman PJ (1988) {Flexibility of DNA}.
\newblock Annual review of biophysics and biophysical chemistry 17: 265--86.
\bibAnnoteFile{Hagerman:1988p595}

\bibitem{Martin:2002p588}
Martin RG, Gillette WK, Martin NI, Rosner JL (2002) {Complex formation between
  activator and RNA polymerase as the basis for transcriptional activation by
  MarA and SoxS in {\it Escherichia coli}}.
\newblock Mol Microbiol 43: 355--70.
\bibAnnoteFile{Martin:2002p588}

\bibitem{de-gennes:1979ij}
De~Gennes PG (1988) Scaling concept in polymer physics.
\newblock Cornell University Press, Ithaca, NY, 3rd edition.
\bibAnnoteFile{de-gennes:1979ij}

\bibitem{Grosberg:1997dc}
Grosberg AY, Khokhlov AR (1997) Statistical Physics of Macromolecules.
\newblock AIP Press.
\bibAnnoteFile{Grosberg:1997dc}

\bibitem{Bloomfield:1997vh}
Bloomfield VA (1997) {DNA condensation by multivalent cations.}
\newblock Biopolymers 44: 269--282.
\bibAnnoteFile{Bloomfield:1997vh}

\bibitem{Englander:2004dq}
Englander J, Klein E, Brumfeld V, Sharma AK, Doherty AJ, et~al. (2004) {DNA
  toroids: framework for DNA repair in {\it Deinococcus radiodurans} and in
  germinating bacterial spores.}
\newblock J Bacteriol 186: 5973--5977.
\bibAnnoteFile{Englander:2004dq}

\bibitem{stukan:2003tg}
Stukan MR, Ivanov VA, Grosberg AY, Paul W, Binder K (2003) Chain length
  dependence of the state diagram of a single stiff-chain macromolecule: Theory
  and monte carlo simulation.
\newblock J Chem Phys 118: 3392-3400.
\bibAnnoteFile{stukan:2003tg}

\bibitem{Langowski:2006p648}
Langowski J (2006) Polymer chain models of dna and chromatin.
\newblock Eur Phys J E 19: 241--9.
\bibAnnoteFile{Langowski:2006p648}

\bibitem{Jackson:1998yw}
Jackson DA, Iborra FJ, Manders EM, Cook PR (1998) {Numbers and organization of
  RNA polymerases, nascent transcripts, and transcription units in HeLa
  nuclei.}
\newblock Mol Biol Cell 9: 1523--1536.
\bibAnnoteFile{Jackson:1998yw}

\bibitem{Vaquerizas:2009p562}
Vaquerizas JM, Kummerfeld SK, Teichmann SA, Luscombe NM (2009) A census of
  human transcription factors: function, expression and evolution.
\newblock Nat Rev Genet 10: 252--63.
\bibAnnoteFile{Vaquerizas:2009p562}

\bibitem{Rosa:2008p511}
Rosa A, Everaers R (2008) Structure and dynamics of interphase chromosomes.
\newblock PLoS Comput Biol 4: e1000153.
\bibAnnoteFile{Rosa:2008p511}

\bibitem{Nicodemi:2009p638}
Nicodemi M, Prisco A (2009) Thermodynamic pathways to genome spatial
  organization in the cell nucleus.
\newblock Biophys J 96: 2168.
\bibAnnoteFile{Nicodemi:2009p638}

\bibitem{Xu:2008dk}
Xu M, Cook PR (2008) Similar active genes cluster in specialized transcription
  factories.
\newblock J Cell Biol 181: 615--623.
\bibAnnoteFile{Xu:2008dk}

\bibitem{Wagner:1999p785}
Wagner A (1999) Genes regulated cooperatively by one or more transcription
  factors and their identification in whole eukaryotic genomes.
\newblock Bioinformatics 15: 776--84.
\bibAnnoteFile{Wagner:1999p785}

\bibitem{Cohen:2000p786}
Cohen BA, Mitra RD, Hughes JD, Church GM (2000) A computational analysis of
  whole-genome expression data reveals chromosomal domains of gene expression.
\newblock Nat Genet 26: 183--6.
\bibAnnoteFile{Cohen:2000p786}

\bibitem{Kosak:2007p607}
Kosak ST, Scalzo D, Alworth SV, Li F, Palmer S, et~al. (2007) Coordinate gene
  regulation during hematopoiesis is related to genomic organization.
\newblock PLoS Biol 5: e309.
\bibAnnoteFile{Kosak:2007p607}

\bibitem{Dostie:2006p561}
Dostie J, Richmond TA, Arnaout RA, Selzer RR, Lee WL, et~al. (2006) {Chromosome
  Conformation Capture Carbon Copy (5C): a massively parallel solution for
  mapping interactions between genomic elements}.
\newblock Genome Res 16: 1299--309.
\bibAnnoteFile{Dostie:2006p561}

\bibitem{Jhunjhunwala:2008p599}
Jhunjhunwala S, van Zelm MC, Peak MM, Cutchin S, Riblet R, et~al. (2008) {The
  3D structure of the immunoglobulin heavy-chain locus: implications for
  long-range genomic interactions}.
\newblock Cell 133: 265--79.
\bibAnnoteFile{Jhunjhunwala:2008p599}

\end{thebibliography}

\section*{Figure Legends}

\begin{figure}[!ht]
\caption{\label{fig:WLC_1type} {\bf 2D cartoon of the three-dimensional self-avoiding WLC model with sparse interacting sites.} 
Sites that can interact are represented by small red filled circle. The outer red circles define the interaction range $d^*$ of the potential. The persistence length $l_p$ ($\sim K/k_B T$) is the typical length beyond which the polymer loses most of its orientational order. See Figure S1 for further details on the polymer description.}
\end{figure}

\begin{figure}[!ht]
\caption{\label{fig:micro_trans} {\bf Sketch of the micro-phase transition
in a single polymer chain.} 
Active interacting sites are indicated by red points. 
From left to right:  Starting from a self-attracting WLC in the swollen state, a sufficient increase of $V_0$, which is indicated by the upward arrow, can lead to the formation of a (compact) homogeneous self-attracting globule. Starting from the latter with a sufficiently high value of $V_0$, a progressive increase  of the distance $\Delta$ between the interacting sites along the polymer (upward arrows) will lead to less compact globules, and eventually to the formation of a micro-structured globule with co-localized sites.}
\end{figure}

\begin{figure}[!ht]
\caption{\label{fig:phasediagprincip} {\bf Macroscopic phase diagram in biologically relevant conditions.} 
Spatial co-localization of co-regulated genes as modeled by a flexible WLC composed of sparse interacting sites, that is having $r_0/\Delta \ll 1$ and $\Delta/l_p \gtrsim 1$. In the case where the attractive interaction of the WLC is not too strong, 
the macroscopic state diagram of the system contains two states (leaving apart the 3-dimensional organization of the foci): 
the micro-structured globule and the swollen state. Fixing $n_I$, the number of sites that belong to the discrete foci in the micro-structured state, the transition lines (dashed red curves) separating the swollen state from the micro-structured globule are of the form $f(x)=n_I^{2/5} x^{1/5}$ -- see relation (\ref{het}). The gray area indicates the typical values taken by $\bar n V_0$ and $\Delta /l_p$ in the eukaryote case. Notice that given an estimation of $10 \lesssim n_I \lesssim 50$ \cite{Jackson:1998yw}, different biologically relevant values of $\bar n V_0$ can allow switching from one state to the other one. }
\end{figure}

\begin{figure}[!ht]
\caption{\label{fig:traveling}{\bf States with micro-structuration of
    interacting sites.}
Parameters correspond to the case of naked DNA with $L=10~\mu{\rm m}$ and $d^*=6~{\rm nm}$.  Left 
panel : $\Delta=l_p$, $V_0 = 2 k_B T$. Right panel: $\Delta=4 l_p$, $V_0 = 5 k_B T$.  Foci are typically composed of 10 interacting sites. Axes give positions in nanometers.}
\end{figure}

\begin{figure}[!ht]
\caption{\label{fig:microphases} {\bf Naive expectation of finite size
    conformations.} When distances (along the one-dimensional DNA) between
  interacting sites are large enough, discrete foci can form in space. This cartoon shows 
  different possible organizations of the foci and of the DNA
  chains. (a) Foci and bundles of DNA free of interacting sites are
  organized along one (thick) dimension. They form either solenoidal
  structures or open linear structures. (b) Foci belong to nodes of a
  spatial network of DNA bundles free of interacting sites. In this
  situation, the DNA chain goes from one focus to another focus that is in its
  spatial vicinity. (c) Foci are organized along one dimensional necklace while DNA chains form rosette structures.}
\end{figure}

\begin{figure}[!ht]
\caption{\label{fig:solenoid_rosette} {\bf Topological ordering of DNA around the foci.} {\it Upper panels:} Conformations
topologically equivalent to solenoids where foci and DNA bundles are
organized in a one-dimensional manner. Naked DNA parameters: $d^*=6~{\rm nm}$, $\Delta=l_p$ and $V_0 = 2 k_B T$. Left panel: $L=0.8\;\mu {\rm m}$; right panel: $L=2.0\;\mu {\rm m}$.   
{\it Lower panels: } Rosettes. Nodes are considered as rosettes
(blue arrows) when more than half of their outgoing DNA chains come back to
the same node at the next interacting site. Chromatin fiber with $l_p=150~{\rm nm}$, $\Delta=4 l_p$. Left panel: for small sizes ($L=6~\mu {\rm m}$), necklaces of no more than two rosettes appear. Right panel: for larger sizes ($L=30~\mu {\rm m}$), foci tend to form random spatial networks instead of long necklaces of rosettes.}
\end{figure}

\begin{figure}[!ht]
\caption{\label{fig:diagramMC} {\bf Qualitative state diagram for finite length WLC.}
Computational tools can provide qualitative insights of the state diagram as a function of the system parameters. 
In this regard, the thick gray lines in the diagram point out the
expected transitions as the parameters are varied -- they do not provide the precise form of the transition lines. The horizontal blue dashed line
is used to simultaneously discuss two limiting cases: very flexible polymers (small $l_p/d^*$) and very rigid polymers (large $l_p/d^*$). 
As far as self-attracting WLC are concerned, at low temperature, the former tend to form spherical globules, whereas the latter
tend to form toroids \cite{stukan:2003tg}.
Now, working with a fixed chain length, our results show that for large enough
values of $\Delta$, the rosette is the most stable state
thermodynamically. Starting from a single rosette, a further
decreasing of $\Delta$ leads to the formation of several foci. These foci can be
organized according to an isotropic spatial network ({\it i.e.},
the traveling chain conformation) as shown in Fig. \ref{fig:traveling}, but also
according to an anisotropic shape as shown in Figure S3. Thus, we go from rosettes to traveling chains when
$\Delta$ decreases, whatever the values of the rigidity.
In this respect, the transition lines separating the rosette and the multi-foci conformations are
expected to lower as the length of the polymer increases (arrows in
the figure). In the same way, as indicated by arrows too, the solenoidal phase decreases in
stability as the length of the polymer increases.
}
\end{figure}

\begin{figure}[!ht]
\caption{\label{fig:rosettevssol} {\bf Impact of genome organization on chromosome structure.} When several types (indicated by different colors of points) of binding sites are present, the specific positioning of
the binding sites has a critical effect on the nature of the chromosomal
structuring. For instance, binding sites that are ordered along the DNA
according to their type, which can be viewed as a clustering of the binding
sites along the DNA, favor the formation of rosettes. This is illustrated in
the left panel (chromatin fiber, $l_p=210$ nm,
$d^*=30$ nm, $\lambda=4l_p$, $V_0=4 k_B T$, $L=34 \mu m$). On the other
hand, a periodic positioning tends to favor a solenoidal organization of the
DNA as illustrated in the right panel (naked DNA. $d^*=6$ nm, $\lambda=2 l_p$, $V_0=3.5 \; k_B T$, $L=4 \mu m$).
}
\end{figure}

\begin{figure}[!ht]
\caption{\label{fig:folding} {\bf Impact of genome organization on the formation of transcription factories.} {\it Left panel:} Folding trajectories traced by the temporal evolution of the total number $N_c$ of contacts that are established among interacting sites. The time is counted in number of sweeps (cf. Methods). The plateau gives the maximum number of contacts. Naked DNA. $N_t=6$, $d^*=6~{\rm nm}$, $L=4~\mu {\rm m}$ and $\lambda=2l_p$. {\it Right panel:} Comparison of the steady state cluster composition between periodic, random periodic and random site positioning.  $c(s)=\rho(s)/s$ is the cluster size distribution. $N_t=8$, $d^*=6~{\rm nm}$, $L=8~\mu {\rm m}$ and $\lambda=2l_p$. See text for the definition of $\rho(s)$.}
\end{figure}

\begin{figure}[!ht]
\caption{\label{fig:orga} {\bf Snapshot of DNA conformations and
    foci within steady states.} Here, foci of maximum
  sizes are reached in all cases. The global conformation depends on 
the way the sites were laid out on the chain. {\it Upper left panel}: random positions; {\it Upper right panel}: periodic positions; {\it Lower panel}: periodic random positions. Naked DNA. $d^*=6$ nm. $V_0=3.5 \; k_B T$. $\lambda=2 l_p$. $L=4\mu m$.}
\end{figure}

\begin{table}[!ht]
    \begin{tabular}[t]{ | c  | p{13.4cm} |}
    \hline
    Parameter & Name and description \\ \hline \hline
    $k_B T$ & {\it Thermal energy:} energy unit reflecting the thermal agitation of the environment ({\it e.g.} the nucleoplasm ) in which the polymer (DNA or chromatin) resides.   \\ \hline
    $l_p$ & {\it Persistence length:} distance beyond which the polymer loses most of its orientational order. \\ & \\
    $K$ & {\it Bending modulus:} energy per unit length. It reflects the energetic cost to locally bend the polymer, which leads to a persistence length $l_p \sim K/k_bT$.  \\ \hline
    $r_0$ & {\it Hard-core radius:} radius of the hard-core polymer. Electrostatic repulsions between DNA segments are therefore modeled as simple hard-core repulsions. \\ & \\
    $R$ & {\it Gyration radius:} spatial extension of a spherical globule conformation of the WLC. $R^2=N^{-2}\sum_{i,j} (\vec r_j- \vec r_i)^2$ where $\vec r_i$ is the position of the $i^{th}$ monomer. \\ \hline
    $V_0$ & {\it Binding free energy:} typical free energy gain due to the bridging of two distal sites along the DNA, not taking account the entropy change of the distant parts of the chain. \\ & \\
    $d^*$ & {\it Interaction range:} Distance below which two interacting sites interact. In such case, they lower the energy of the system by an amount of $-V_0$ ($V_0>0$).\\ \hline
    $\Delta$ &  Mean distance between two successive interacting sites along the DNA. \\ & \\
    $\bar n$ &  Maximum number of partners of a single interacting site. Biologically speaking, this can be viewed as the maximum number of TF binding sites of a regulated gene. \\ \hline
    $n_I$ & Typical number of genes that are transcribed simultaneously in a transcription factory (biological data). This corresponds, here, to the mean number of interacting sites belonging to the discrete foci.\\
    \hline
    $\lambda$ & Mean distance between two consecutive sites. For one type of interacting sites, $\lambda=\Delta$.\\
    \hline
    \end{tabular}
        \caption{List of parameters. \label{table}}
\end{table}

\end{document}